# The role of the false events on the DQE measurement of the radiation detectors

## Giovanni Zanella

*Dipartimento di Fisica e Astronomia "Galileo Galilei", Università di Padova and*

*I.N.F.N.-Sezione di Padova, via Marzolo, 8 -I-35131 Padova, Italy*

---

**Abstract**

 The efficiency of a radiation detector, intended as probability of detection of an incident quantum, depends on various factors: the detected fraction of quanta ascribed to the noise-less detector, the intrinsic noise of the detector, the false events introduced by the detection process and the false events generated by the detector itself.
 In this paper is treated the role of the false events on the measurement of the detective quantum efficiency (DQE) of the radiation detectors, indifferently for counting detectors and imaging detectors.

---

## 1. Introduction

 In general, we can define *detection efficiency* of a radiation detector the *probability $\varepsilon$ to detect an incident quantum on the detector itself*, that is the ratio between the *average number of the detected quanta* at the detector output and the *average number of the incident quanta* on the detector [1], so

$$\varepsilon = <detected\ quanta>/<incident\ quanta> . \qquad (1)$$

 Now, *false* (or spurious) *events*, which are indistinguishable from the *detected quanta*, can be included in the measurement of *<detected quanta>*. Obviously, the presence of *false events* depresses the measured *detection efficiency* of the detector, while, on the contrary, the quantity *<detected quanta>* increases, due to impossibility to distinguish *false events* from *detected quanta*.

Hence, the expression (1) has a theoretical validity, but it is not useful for the measurement of $\varepsilon$.

We can define two types of *detection efficiency*: the *quantum efficiency* (*QE*), which supposes the absence of *false events* in the noise-less detector and the *detective quantum efficiency* (*DQE*) which including the *false events* in the noisy detector, becomes the real *detection efficiency*.

As concerns the measurement of *DQE*, we have considered in a previous paper [2] only the incidence of the *false events* introduced by the detector. Instead, in the present paper, we consider two types of *false events* :

1. *False events* generated by the detector and as such their number is averagely steady during a same time interval and on a same detector area.

2. *False events* generated in the interaction of the *incident quanta* with the detector (i.e. Compton events in a gamma detector) and as such their number increases with the *incident quanta*.

Now, if the incidence on *DQE* of the *false events* of type 1 can be nullified, for the Poisson fluctuations associated to an elevated number of *incident quanta*, the *false events* of type 2 result more dangerous,  because they depress *DQE* without remedy.

Due to impossibility to distinguish *detected quanta* from *false events*, the *false events* of type 2 influence the measurement of *QE*, forcing us to its computation, necessary for the determination of *DQE*.

The following treatment is common to the *counting detectors* (including the visualizing ones), in which the events are recognized singularly, and to the *integrating detectors*, in which the events are summed in the pixels of the detector to form images.

Besides, while the output signal is intended on all the bandwidth of the detector.

## 2. The quantum efficiency (*QE*)

Applying Eq. (1) to a noise-less detector and in absence of *false events*, we can write

$$QE = \frac{\overline{S_o}}{\overline{S_i}} \quad , \qquad\qquad (2)$$

where $\overline{S_o}$ denotes the average number of *detected quanta* at the detector output and $\overline{S_i}$ the average number of *incident quanta* on the detector.

Obviously, the previous means are intended on a same interval and on a same area of detector.

It is possible to find an equivalent representation of *QE*, in place of Eq.(2), if Poisson statistics is attributed to the *incident quanta.* Indeed, $\overline{S_i}$ represents the variance $\sigma_i^2$ of the *incident quanta*, while the variance of the *detected quanta* is $\sigma_o^2 = QE\overline{S_i}$ [3]. Thus

$$QE = \frac{\overline{S_o}}{\overline{S_i}} = \frac{\dfrac{\overline{S_o}^2}{\overline{S_o}}}{\dfrac{\overline{S_i}^2}{\overline{S_i}}} = \frac{\dfrac{\overline{S_o}^2}{\sigma_o^2}}{\dfrac{\overline{S_i}^2}{\sigma_i^2}} = \frac{\overline{SNR_o}^2}{\overline{SNR_i}^2},$$ (3)

where $\overline{SNR_o}$ denotes the average signal-to-noise ratio at the detector output and $\overline{SNR_i}$ the average signal-to-noise ratio at the detector inpu.

As, the measurement of *QE* is not possible in presence of *false events*, for the indistinguishability with the *detected quanta*, it is necessary its indirect determination, either as limit case of *DQE*, when we increase $\overline{S_i}$, or by its computation, if nature and energy of the incident particles, as well as the interaction process with the detector, are known.

## 3. *DQE* and false events

In analogy with Eq.(3), *DQE* of a radiation detector is commonly defined as:

$$DQE = \frac{\overline{SNR_o}^2}{\overline{SNR_i}^2} = \frac{\dfrac{\overline{S_o}^2}{\sigma_o^2}}{\dfrac{\overline{S_i}^2}{\overline{S_i}}} = \frac{QE^2\,\overline{S_i}^2}{\overline{S_i}\,\sigma_o^2} = \frac{QE^2\,\overline{S_i}}{\sigma_o^2},$$ (4)

where $\sigma_o^2$ is not simply the Poisson variance $QE\overline{S_i}$ of Eq.(3), because the presence of *false events* and the intrinsic noise of the detector is now considered. Instead, $\overline{S_o}^2$ of Eq.(4) is necessarily $QE^2\overline{S_i}^2$ as in Eq.(3), because the *false events* must be excluded from the output signal, which pertains only the *detected quanta*, while other Gaussian noises do not alter its average value.

As concerns $\sigma_o^2$ of Eq.(4), it is the sum of various independent variances, that is (ascribing Poisson statistics also to the *false events*)

$$\sigma_0^2 = QE\overline{S_i} + \overline{f}\,QE\overline{S_i} + \overline{f_d} + \sigma_d^2 \quad , \qquad\qquad (5)$$

where: $\overline{f}$ is the *average number of false events per detected quantum*, $\overline{f_d}$ is the *average number of false events generated by the detector* and $\sigma_d^2$ represents the variance of the Gaussian noise introduced by the detector.

In Eq.(5) $\overline{f}$ and $\overline{f_d}$ are obviously measured on the same time interval and detector area of $\overline{S_i}$ and $\overline{S_o}$. Therefore, we can write

$$DQE = \frac{QE}{1 + \overline{f} + \dfrac{\overline{f_d} + \sigma_d^2}{QE\,\overline{S_i}}} \quad , \qquad\qquad (6)$$

It is evident in Eq.(6) that if $\overline{f} = \overline{f_d} = \sigma_d^2 = 0$ (absence of *false events* in a noise-less detector) thus $DQE = QE$, so $DQE \leq QE$ in any case.

The contribution of $\overline{f}$ appears in the following simple example: if the number of the *false events* is averagely equal to the number of the *detected quanta*, that is $\overline{f} = 1$, so the *detection efficiency* is halved, also if $\overline{f_d} = \sigma_d^2 = 0$.

It is evident in Eq. (6) that, increasing $\overline{S_i}$, the noise variance $QE\overline{S_i} + \overline{f}\,QE\overline{S_i}$ dominates on $\overline{f_d} + \sigma_d^2$, so $DQE$ can tend to $QE/\left(1 + \overline{f}\right)$, or to $QE$, if $\overline{f} = 0$.

Therefore, differently on $QE$, $DQE$ depends on $\overline{S_i}$, as well as on the nature of the *incident quanta* and on the detector.

## 4. *DQE* measurement

We can rewrite Eq.(6) as

$$DQE = \frac{QE^2\,\overline{S_i}}{QE\overline{S_i} + \overline{f}\,QE\overline{S_i} + \overline{f_d} + \sigma_d^2} = \frac{QE^2\,\overline{S_i}}{\sigma_n^2} \qquad (7)$$

where $\sigma_n^2$ is the total variance of the signal at the detector output. So, the measurement of *DQE* involves measurable quantities as $\overline{S_i}$ and $\sigma_n^2$, but also a not-measurable quantity in presence of *false events*, that is *QE*. On the contrary, looking at Eq.(6), if the *false events* are due only to detector, or they do not appear, it would be possible to measure *QE* as limit of by *DQE* increasing $\overline{S_i}$ .

Therefore, when there are *false events* associated to the *detected quanta* ( $\overline{f} \neq 0$ ), *QE* must be computed by knowing nature and energy of *incident quanta*, as well as the physical properties of the detector and the interaction particle-detector. Hence, we can determine *DQE*, being known *QE*, by measuring (in quanta) both $\overline{S_i}$ , with an independent detector, and $\sigma_n^2$ by the fluctuations of $\overline{S_o}$ .

Attention must be paid to the measurement of $\sigma_n$ , because if $\overline{S_i}$ is measured in quanta, also $\sigma_n^2$ must be measured in quanta. Thus, if the detector is not counting, it is necessary to know the detector response (pixel by pixel) to a single *incident quantum* and to distribute averagely this response on the same detector area, involved in the measurement of the fluctuations of $\overline{S_o}$ in quanta.

## 5. Conclusions

This paper treats the role of the *false events* in the measurement of *QE* and *DQE* of any radiation detector. Two types of *false events* are involved in the detection process:
1. *False events* which depend only from the detector.
2. *False events* associated to the *detected quanta*.

While the incidence on *DQE* of the *false events* of type 1 can be vanished by increasing the mean of the *incident quanta*, *false events* of type 2 engrave on *DQE* as $QE/(1+\overline{f})$ , being $\overline{f}$ the mean of *false events* per *detected quantum*. Besides, events of type 2 prevent the measurement of *QE*, which . knowledge is essential for the determination of *DQE*, forcing us to its *12*. analytical calculation.

In this paper, the treatment has been common both to the counting detectors (including the visualizing ones) and the integrating (also imaging) detectors, while the output signal is intended on all the bandwidth of the detector.

## 6. References

[1] G. F. Knoll, Radiation Detection and Measurement J. Wiley & Sons, Inc. N.Y. (2000).
[2] G. Zanella, Nucl. Instr. and Meth. A 586 (2008) 372-373
[3] G. Zanella and R. Zannoni, Nucl. Instr. and Meth. A 381 (1996) 157-160.